\documentclass[reprint,amsmath, superscriptaddress, amssymb, aps, longbibliography, prl]{revtex4-1}
\usepackage{graphicx}
\usepackage{dcolumn}
\usepackage{bm}
\usepackage[T1]{fontenc}
\usepackage[utf8]{inputenc}
\usepackage{physics}
\usepackage{xcolor}
\newcommand\hl[1]{%
  \bgroup
  \hskip0pt\color{blue}%
  #1%
  \egroup
}
\DeclareMathAlphabet{\pazocal}{OMS}{zplm}{m}{n}

\begin{document}
\linespread{1}

\title{Polaron-induced changes in moiré exciton propagation in twisted van der Waals heterostructures}
\author{Willy Knorr}
\email{knorrw@uni-marburg.de}
\author{Samuel Brem}
\author{Giuseppe Meneghini}
\author{Ermin Malic}
\affiliation{Department of Physics, Philipps University, 35037 Marburg, Germany}
\begin{abstract}
    Twisted transition metal dichalcogenides (TMDs) present an intriguing platform for exploring excitons and their transport properties. By introducing a twist angle, a moiré superlattice forms, providing a spatially dependent exciton energy landscape. Based on a microscopic many-particle theory, we investigate in this work polaron-induced changes in exciton transport properties in the exemplary MoSe$_2$/WSe$_2$ heterostructure. We demonstrate that polaron formation and the associated enhancement of the moiré exciton mass lead to a significant band flattening. As a result, the hopping rate and the propagation velocity undergo noticeable temperature and twist-angle dependent changes. We predict a reduction of the hopping strength  ranging from 80\% at a twist angle of 1$^\circ$ to 30\% at 3$^\circ$ at room temperature. The provided  microscopic insights into the spatio-temporal exciton dynamics in presence of a moiré potential further expand the possibilities to tune charge and energy transport in 2D materials.\\    
\end{abstract}
\maketitle
\section{Introduction}
Excitons, fundamental elements of condensed matter physics, play a critical role in determining optics, dynamics, and transport properties of  transition metal dichalcogenides (TMDs) \cite{wang2018colloquium, mueller2018exciton,perea2022exciton,rosati2021electron}.  TMDs belong to the family of truly two-dimensional nanomaterials and have recently emerged as a fascinating platform for studying exciton physics and in particular their propagation behaviour at room temperature. Interesting effects range from tunable exciton transport in TMD monolayers subject to strain and dielectric engineering \cite{niehues2018strain, aslan2018strain, latini2015excitons, rosati2021dark,khatibi2018impact,feierabend2017impact} including non-linear effects, such as the formation of spatial rings (halos) \cite{kulig2018exciton, perea2019exciton}, as well as non-classical exciton diffusion \cite{wagner2021nonclassical}, and even effective negative diffusion \cite{rosati2020negative, berghuis2021effective,wietek2023non}. When TMD monolayers are vertically stacked into van der Waals heterostructures, they exhibit intriguing many-body phenomena, offering rich exciton energy landscapes that can be tuned with the twist angle \cite{yu2017moire, merkl2019ultrafast, brem2020tunable, shabani2021deep, merkl2020twist,schmitt2022formation,meneghini2022ultrafast}.
 \begin{figure}[t!]
    \includegraphics[width=0.49\textwidth]{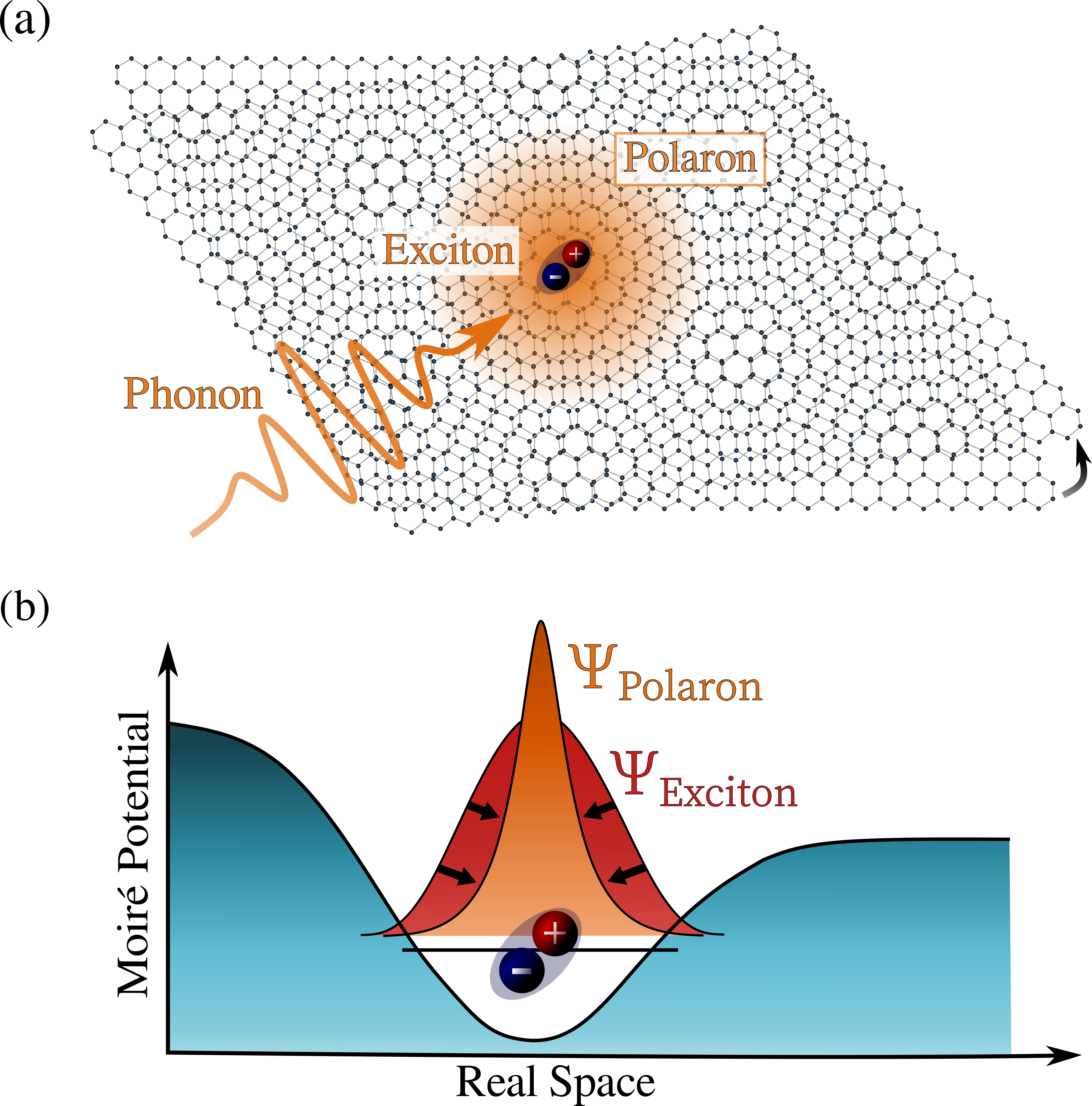}
    \caption{(a) Schematic on the coupling between excitons and lattice vibrations (phonons), resulting in the formation of polarons in a twisted TMD heterostructure. (b) Schematic illustration of the polaron-induced spatial narrowing of the excitonic wave function in presence of a moiré potential. This effect is temperature and twist-angle dependent and has a direct impact on the exciton band width and the exciton propagation behaviour.}
    \label{fig:Schema}
\end{figure}

In a recent study, we have shown a theoretical framework on moiré excitons including their spatio-temporal dynamics in van der Waals heterostructures \cite{knorr2022exciton}. This allowed us to study different twist-angle dependent moiré exciton transport regimes in an hBN-encapsulated MoSe$_2$/WSe$_2$ heterostructure. 
However, the important role of polarons has not been addressed in this context. Polarons have recently moved in the focus of research on twisted TMD heterostructures \cite{mazza2022strongly, huang2023mott, campbell2022exciton}. 
They are a result of an efficient coupling of excitons and lattice vibrations (phonons), cf. Fig. \ref{fig:Schema}.
As TMDs are known to exhibit an efficient exciton-phonon interaction \cite{selig2016excitonic,schmitt2022formation,brem2020phonon,brem2018exciton}, polarons are expected to play an important role also for exciton transport in TMDs. 
In particular, it is an interesting question to investigate how polaronic effects change the band structure and the propagation behaviour of moiré excitons, which can become trapped, i.e. exhibit flat bands, at small twist angles. 
Thus, understanding the impact of the polaron formation on moiré exciton transport is highly interesting and may even have implications on emergent phenomena, such as superconductivity, magnetism, and topological phases \cite{alexandrov1992polarons,mauger1983magnetic,qin2019polaron}. Moreover, the influence of polarons on charge transport, optical absorption spectra, and exciton dynamics could enable new pathways for technological applications of TMD materials \cite{vandewal2017absorption,van1990polaronic,johansson2004nonadiabatic}.

In this work, we theoretically investigate  the impact of polarons on the moiré exciton band structure and the propagation behaviour in the twisted  MoSe$_2$/WSe$_2$ heterostructure. Based on a microscopic theory we determine the change in the excitonic eigenstates within a moiré potential and calculate the twist-angle and temperature dependent hopping rates governing the propagation of moiré excitons. We find that polarons considerably enhance the mass of moiré excitons and contribute to a band flattening. This has an immediate impact on moiré exciton hopping rates and their propagation velocity. We show a distinct temperature dependence and reveal a considerable slow-down of the exciton propagation in the range of small twist angles.

\section{Theory}
In the following, we set up a microscopic theory allowing us to study the propagation of excitons within a moiré potential taking explicitly into account the formation of polarons. 
We start with an excitonic Hamilton operator in the low-density regime and find an exact numerical solution for the scattering-free propagation of moiré excitons. For large twist angles, the moiré mini-band structure converges into the free exciton dispersion, such that moiré excitons are expected to propagate like quasi-free particles \cite{knorr2022exciton}. On the other hand, the flat band structure at small twist angles indicates a strongly reduced exciton propagation, suggesting hopping-dominated transport \cite{knorr2022exciton}. We expand our theoretical framework by transforming the exciton Hamiltonian into a polaron basis, which takes into account the interaction with phonons. This allows us to explore the influence of exciton-phonon interactions on the transport behavior of excitons within the exemplary hBN-encapsulated MoSe$_2$/WSe$_2$ heterostructure.

\textbf{Moiré Excitons.}
Twisting atomically thin TMD monolayers within a heterostructure leads to the formation of a moiré superlattice introducing a spatially dependent exciton energy landscape. To study this, we express the Hamilton operator in an exciton basis \cite{haug1984electron,katsch2018theory} 
\begin{align}\label{Eq:singleHamiltonian_exciton}
    H=\sum_{\mu\mathbf{Q}}\mathcal{E}_{\mathbf{Q}}^{\mu}X^{\dagger}_{\mu\mathbf{Q}}X_{\mu\mathbf{Q}}+\sum_{\mu\mathbf{Q}\mathbf{q}}\mathcal{M}_{\mathbf{q}}^{\mu}X^{\dagger}_{\mu\mathbf{Q}+\mathbf{q}}X_{\mu\mathbf{Q}}
\end{align}
using excitonic creation and annihilation operators $X^{(\dagger)}_{\mu\mathbf{Q}}$ \cite{brem2020tunable}. The first part of the Hamiltonian represents the free contribution with the exciton dispersion $\mathcal{E}_{\mathbf{Q}}^{\mu}$, while the second part involves the moiré potential with the matrix element $\mathcal{M}_{\mathbf{q}}^{\mu}$ (see SI for more details). Here, $\mathbf{Q}$ represents the center-of-mass momentum, while $\mu$ corresponds to different exciton states including intra- and interlayer excitons \cite{merkl2019ultrafast,schmitt2022formation}. In the low-density regime, exciton-exciton interactions can be disregarded \cite{erkensten2021exciton,brem2023bosonic}, which simplifies the analysis. Exciton-phonon coupling will be explicitly taken into account in the next section. 

For the specific case of the MoSe$2$/WSe$2$ heterostructure investigated here, KK interlayer excitons are known to be  the lowest energy states \cite{lu2019modulated, brem2020tunable} and the hybridization between intra- and interlayer exciton states can be neglected \cite{gillen2018interlayer, brem2020hybridized}. By employing a zone-folding approach \cite{brem2020tunable, brem2020hybridized, brem2022terahertz, brem2023bosonic}, exploiting the moiré potential's periodicity, we map points outside of the mini-Brillouin zone back inside, using the reciprocal lattice vectors $\mathbf{G}$ of the superlattice. The resulting eigenvalue problem is local in momentum space and corresponds to a mixing of discrete dispersion branches. This results in a band structure represented by a series of moiré exciton subbands $\nu$. In this moiré-exciton basis, the Hamiltonian becomes diagonal reading 
\begin{align}\label{Eq_single_particle_final}
    H = \sum_{\nu\mathbf{Q}}E_{\mathbf{Q}}^{\nu}Y^{\dagger\nu}_{\mathbf{Q}}Y^{\nu}_{\mathbf{Q}}, \quad Y^{\dagger\nu}_{\mathbf{Q}} = \sum_sc^{*\nu}_{s}(\mathbf{Q})X^{\dagger}_{\mathbf{Q}+\mathbf{G}_s}
\end{align}
where  $E_{\mathbf{Q}}^{\nu}$ and  $c^{*\nu}_{s}(\mathbf{Q})$ are energy eigenvalues and  eigenstates, respectively. 
Further details can be found in the supplementary material.

After determining the excitonic eigenstates within the moiré potential, we proceed with computing the time- and space-dependent wave functions, starting from a given initial distribution $\tilde{\psi}_{\nu}(\mathbf{Q},0)$, where we  populate only the lowest moiré subband. This inital state is represented by a superposition of Bloch wave functions $\chi_{\mathbf{Q},\nu}(\mathbf{r})$ weighted by a Gaussian distribution reminiscent of a laser pulse. This results in a broadly dispersed initial wave function across the real space. The evolution of this wavefunction over time is governed by the  time evolution operator $U(t) = \text{exp}(-\frac{i}{\hbar}E_{\mathbf{Q}}^{\nu}t)$ resulting in the following equation for the moiré exciton wavefunction
\begin{align}\label{Eq:exact_solution}
    \psi(\mathbf{r},t)=\sum_{\mathbf{Q},\nu}\tilde{\psi}_{\nu}(\mathbf{Q},0)\chi_{\mathbf{Q},\nu}(\mathbf{r})\,\text{exp}\Big(-\frac{i}{\hbar}E_{\mathbf{Q}}^{\nu}t\Big).
\end{align}
Now, we are able to compute the corresponding exciton distribution, allowing us to determine the variance $\sigma^2(t)=\int d^2r\mathbf{r}^2\rho(\mathbf{r},t)$. This describes the broadening of the initial distribution over time. Through the variance we can define a dispersion length $\xi(t) = \sqrt{\sigma^2(t) - \sigma_0^2}$, where $\sigma_0^2$ denotes the variance of the initial density distribution \cite{yuan2020twist}. To obtain access to the propagation velocity of moir\'e excitons, we now define a dispersion velocity $\alpha$, characterizing the temporal broadening
\begin{align}\label{Eq:alpha}
    \alpha = \frac{\sigma_0^2}{2}\partial_t \xi(t).
\end{align}
In the case of a parabolic dispersion, it is valid that $\alpha = \hbar/m$. 

In this study, we explore the moiré exciton transport, considering scenarios where the exciton band structure ranges from being flat to exhibiting strongly dispersed bands \cite{brem2020tunable}. To model this, we adopt a moiré inter-cell tunneling approach, successfully used in describing bosonic atoms in optical lattices \cite{jaksch1998cold, jaksch2005cold} and quantum phases transitions \cite{jaksch1998cold, greiner2002quantum}.
Starting with the exciton Hamiltonian, Eq.\eqref{Eq_single_particle_final}, we transform the operators into a Wannier basis $b^{\dagger}_{\nu,n} = 1/\sqrt{N}\sum_{\mathbf{Q}}\text{exp}(i\mathbf{Q}\cdot\mathbf{R}_n)Y^{\nu\dagger}_{\mathbf{Q}}$ representing strongly localized hopping-driven states. The resulting Hamiltonian reads
\begin{equation}\label{Eq:Hopping_Hamiltonian}
    H = \sum_{n,m,\nu} t_{n,m}^{\nu} b^{\dagger}_{\nu,n}b_{\nu,m}
\end{equation}
with the hopping strength $t_{n,m}^{\nu}$. The latter depends on the overlap of the Wannier wavefunctions $W_{n}(\mathbf{r})=1/\sqrt{N}\sum_{\mathbf{Q},s}e^{-i\mathbf{Q}\cdot\mathbf{R}_n}c_s^{\nu}(\mathbf{Q})\text{exp}\{-i(\mathbf{Q}+\mathbf{G}_s)\mathbf{r}\}$ at lattice positions $n$ and $m$ resulting in
\begingroup
\begin{align}\label{Eq:Hopping_term}
    t^{\nu}_{n,m} &=\frac{1}{N}\sum_{\mathbf{Q}}e^{i\mathbf{Q}\cdot(\mathbf{R}_m-\mathbf{R}_n)}E_{\mathbf{Q}}^{\nu}
\end{align}
\endgroup
with the twist-angle-dependent moiré exciton band structure $E_{\mathbf{Q}}^{\nu}$.
 Flat bands with localized states lead to small hopping terms due to a limited Wannier orbital overlap, while parabolic bands with delocalized states have larger hopping contributions. These hopping terms provide a comprehensive representation of the moiré eigenstates in the Wannier basis. Our study focuses on the lowest orbital and considers Hubbard-like tunneling between nearest neighbors for simplicity.

\textbf{Polaron Formation.}
Now, we introduce the exciton-phonon Hamiltonian that has been disregarded so far. Given our emphasis on interlayer excitons in the K-valley, specifically targeting the lowest subband, we omit the subband index $\nu$. Additionally, for the sake of simplicity, we omit the phonon mode index $j$, although it remains relevant, even though it is not explicitly mentioned. Consequently, we can simplify the Hamiltonian yielding \cite{brem2020hybridized,wallauer2021momentum}
\begin{align}
    H_{x-ph} &=  \sum_{\mathbf{Q},\mathbf{q}}D_{\mathbf{q}}X_{\mathbf{Q}+\mathbf{q}}^{\dagger}X_{\mathbf{Q}}(b_{\mathbf{q}}+b^{\dagger}_{-\mathbf{q}}).
\end{align}
Here, $\mathbf{Q}$ represents the center-of-mass momentum of the exciton, while $\mathbf{q}$ denotes the momentum transfer due to exciton-phonon interaction. Furthermore, $D_{\mathbf{q}}$ denotes the exciton-phonon matrix element, which describes the interaction strength between the exciton and a phonon with the momentum $\mathbf{q}$. This coupling strength is derived by transforming the corresponding electron-phonon matrix element extracted from DFT calculations \cite{jin2014intrinsic} into the exciton basis (see SI for more details).

We now perform a canonical transformation (Lang-Firsov transformation \cite{lang1963kinetic}),  where we introduce a transformation operator that decouples electrons from lattice vibrations. This transformation makes the system's behavior more intuitive, aiding in the analysis of phenomena like polaron formation and other electron-phonon coupling effects. This leads us to the following Hamiltonian
\begin{align}
    \widetilde{H} &= H - \sum_{\mathbf{Q},\mathbf{q}}|D_{\mathbf{q}}|^2\left(\frac{n_q}{\Delta \mathcal{E}-\omega_{\mathbf{q}}}+\frac{n_{\mathbf{q}}+1}{\Delta \mathcal{E}+\omega_{\mathbf{q}}}\right)X_{\mathbf{Q}}^{\dagger}X_{\mathbf{Q}}
\end{align}
with $n_{\mathbf{q}}$ representing the phonon occupation number  and $\Delta \mathcal{E} =\mathcal{E}_{\mathbf{Q}+\mathbf{q}}-\mathcal{E}_{\mathbf{Q}}$. This Hamiltonian gives rise to the  polaron-renormalized energy bands $\widetilde{\mathcal{E}}_\mathbf{Q}$. Carrying out a Taylor expansion we find
\begin{align}\label{Eq:Polaron_Energies}
    \widetilde{\mathcal{E}}_\mathbf{Q} &= (1 -\lambda) \mathcal{E}_{\mathbf{Q}} -\mathcal{E}_{\text{Polaron}} = \frac{\hbar^2\mathbf{Q}^2}{2m^*} -\mathcal{E}_{\text{Polaron}} .
\end{align}
Here, $\mathcal{E}_{\text{Polaron}} = \frac{1}{N} \sum_q \frac{|D_{\mathbf{q}}|^2(n_{\mathbf{q}}+1)}{\mathcal{E}_{\mathbf{q}}+\hbar\Omega}$ represents a polaron-induced shift in energy and $\lambda = \frac{2\hbar^2}{N m} \sum_{q} \frac{|D_{\mathbf{q}}|^2 \mathbf{q}^2 (n_{\mathbf{q}}+1)}{(\mathcal{E}_{\mathbf{q}}+\hbar\Omega)^3}$ denotes a change in the excitonic mass $m$ to a polaron mass $m^* = (1+\lambda)m$. In our model, we only include the coupling to optical phonons $\omega_{\mathbf{q}} = \hbar\Omega$, since the continuous low energy spectrum of acoustic phonons gives rise to short phonon coherence times, which suppresses the coherent hybridization of excitons and phonons.  By incorporating the polaron-renormalized energies into Eq.\eqref{Eq:singleHamiltonian_exciton}, we can now delve into a detailed examination of how polaronic mass-enhancements affect the moiré exciton band structure. Additionally, by utilizing the obtained polaronic band structure and applying them to Eq.\eqref{Eq:exact_solution}, we can investigate how the formation of polarons influences the spatial propagation of moiré excitons in twisted TMD heterostructures.

\section{Exciton Polaron Transport}
\begin{figure}[t!]
    \includegraphics[width=0.48\textwidth]{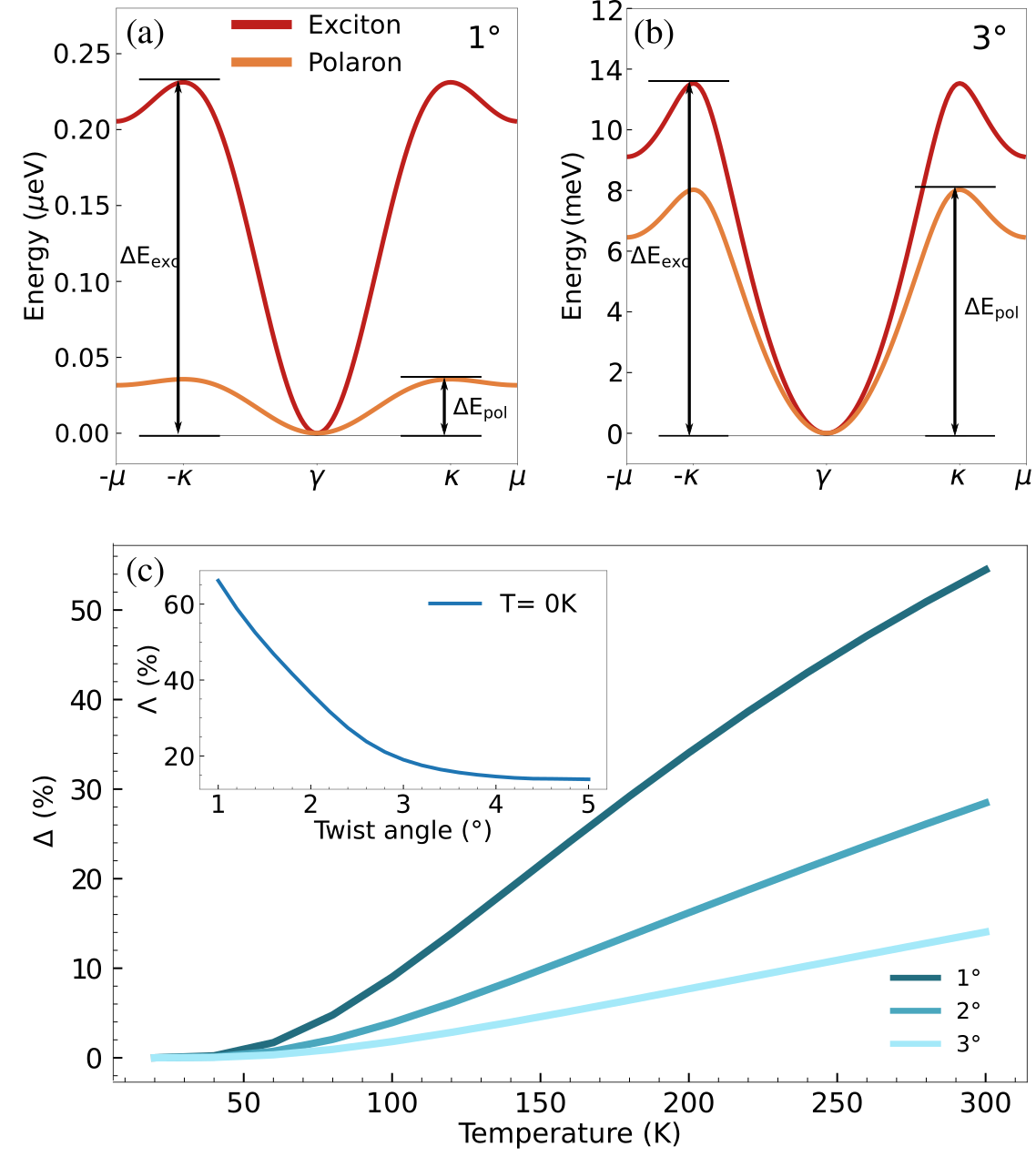}
    \caption{(a)-(b) Comparison of the lowest moiré exciton subband structure of the twisted MoSe$_2$/WSe$_2$ heterostructure including the formation of polarons (orange) with the purely excitonic case without the interaction with phonons (red), renormalized to their respective minima. The effect is measured by studying the change in the exciton bandwidth ($\Delta\text{E}_{\text{exc}}$) and the polaron bandwidth ($\Delta\text{E}_{\text{pol}}$). (c) The relative polaron bandwidth change $\Delta=1-\Delta\text{E}_{\text{pol}}(T)/\Delta\text{E}_{\text{pol}}(0)$ as a function of temperature for different twist angles (with $\Delta\text{E}_{\text{pol}}(T=0)$ as reference). We find a significant change in the polaron bandwidth of over 50\% for a twist angle of 1$^\circ$. This is further illustrated in the inset, where we show the relative twist-angle dependent change in the polaron bandwidth  $\Lambda=1-\Delta\text{E}_{\text{pol}}(0)/\Delta\text{E}_{\text{exc}}(0)$ with respect to the exciton bandwith at a fixed temperature of T=0 K.}
    \label{fig:Bandwidth}
\end{figure}

In recent years, the study of polaron effects in two-dimensional materials, particularly in TMDs, has attracted significant attention due to its potential in tailoring material properties \cite{franchini2021polarons,singh2022electric,rodin2020collective}. One important polaron effect is the emergence of renormalized energy bands. In the following, we characterize moir\'e mini-bands by their bandwidth $\Delta E$, which corresponds to the spectral range between the maximum and minimum energy  within the band (Fig.\ref{fig:Bandwidth} (a)). 

\textbf{Polaron impact on moiré exciton bandwidth.} First, we study the twist-angle dependent influence of the polaron-induced mass enhancement on the moiré exciton bandstructure. We find a distinct flattening of the bands, i.e. a notable reduction in the bandwidth, cf. Figs. \ref{fig:Bandwidth}(a)-(b). The absolute change in the bandwidth decreases by approximately 6 meV at large twist angles (3$^{\circ}$), while smaller twist angles (1$^{\circ}$) exhibit changes of only 20 $\mu$eV, attributed to their inherently flatter bands. We explore the influence of temperature on the polaron-induced bandwidth change, where we calculate the relative change in the bandwidth $\Delta=1-\Delta E_{\text{pol}}(T)/\Delta E_{\text{pol}}(0)$, i.e. taking the polaronic bandwidth $\Delta E_{\text{pol}}(0)$ at T=0K as reference (see Fig. \ref{fig:Bandwidth}(c)). We observe an interesting temperature-dependent behavior. This originates from the increased phonon occupation numbers $n_{\mathbf{q}}$ at elevated temperatures. Thus, we obtain an increase in the effective excitonic mass, as elucidated by Eq \eqref{Eq:Polaron_Energies}, consequently fostering a recognizable flattening of the band structure (see also Fig. \ref{fig:Bandwidth}(a)-(b)).  We find the largest band width change at large temperatures and small twist angles, leading to a 50\% increase at  1$^{\circ}$ and at room temperature. In contrast, at 3$^{\circ}$, the relative change is much smaller with approximately 10\%. Although the polaron formation is not directly twist-angle dependent, the interplay between the polaron and the moiré potential leads to a significant twist-angle dependence in the bandwidth. This behavior results from the fact that small changes in band flattening, due to polaron formation, lead to a large relative change in already very flat bands. On the other hand, at larger bandwidths, small changes in the band structure lead to less pronounced effects. Nevertheless, these  observations show us that small twist angles are more strongly influenced by the formation of polarons. This becomes even more evident if we consider the relative change in the band structure compared to the unperturbed exciton ($\Lambda = 1-\Delta E_{\text{pol}}/\Delta E_{\text{exc}}$), cf. the inset of Fig. \ref{fig:Bandwidth}(c). The change in the band width is particularly large at small twist angles, exhibiting  remarkable relative changes exceeding 60\% (at T = 0K and at  1$^{\circ}$). As the twist angle increases, the magnitude of this effect diminishes significantly, with differences falling below 20\%. \\
While in general, the polaronic mass-enhancement is independent of the twist angle, its impact on the moir\'e trapping very strongly depends on the length scale of the trapping potential. The delocalization of excitons at large twist angles observed in previous works \cite{brem2020tunable, knorr2022exciton} is a direct result of the zero-point energy of quantum confined states. When decreasing the confinement length scale (increasing the twist angle), the zero-point energy of the ground state exciton becomes increased. At some critical confinement length the zero-point energy is larger then the depth of the moir\'e potential, which leads to a delocalization of the exciton accompanied by increase in the band width. However, the zero-point energy of a quantum confined state is inversely proportional to the particle mass. Therefore, the polaronic mass enhancement has a different impact on the moir\'e bandwidth in the trapped regime (small twist angles) than in the delocalized regime (large twist angles).  
Overall, our finding illustrate a nuanced relationship between the twist angle, temperature, and the extent of bandwidth modifications induced by polarons.

\begin{figure}[t!]
    \centering
    \includegraphics[width=0.48\textwidth]{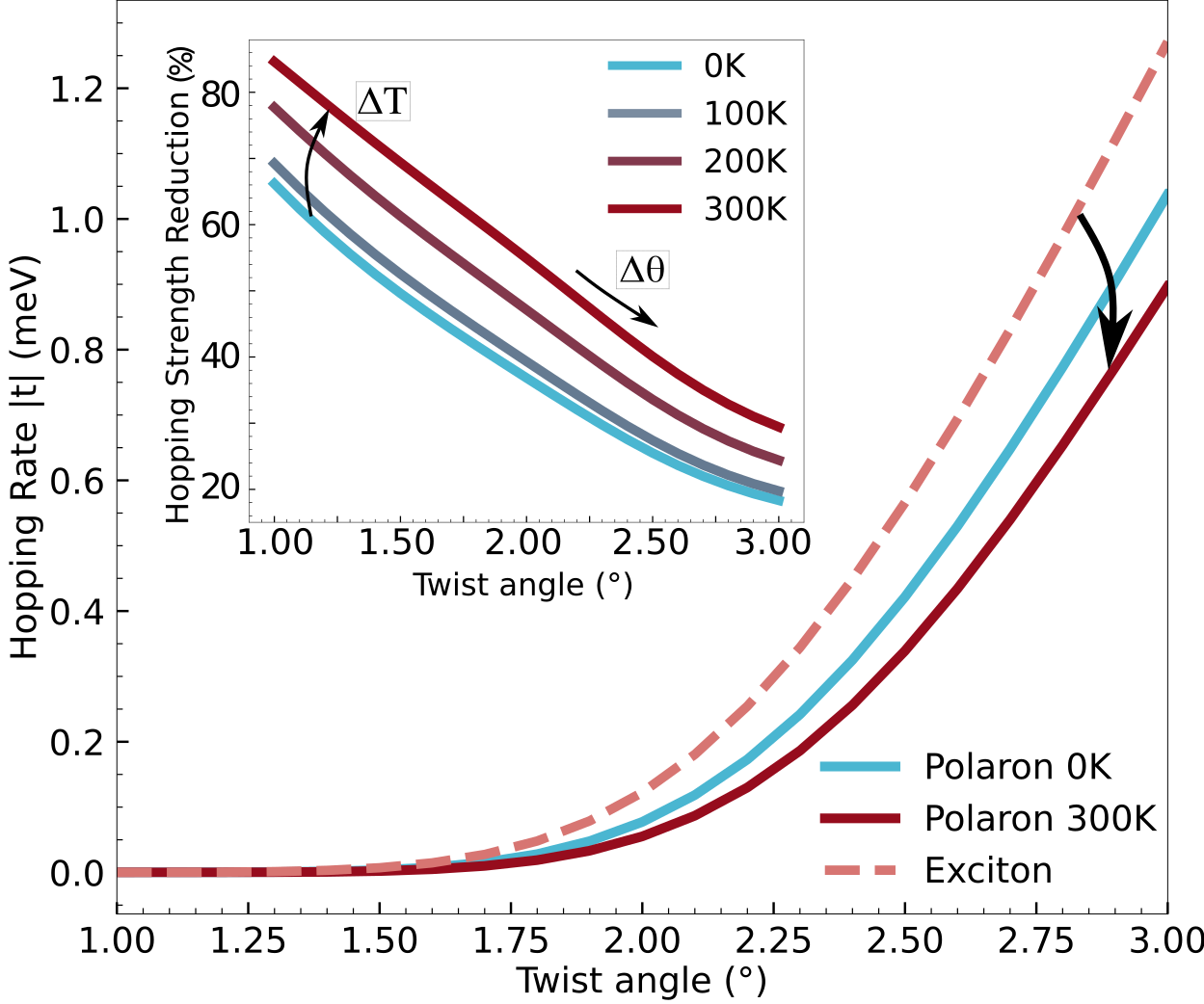}
    \caption{The hopping rate $|t|$ as a function of the twist angle with and without the presence of polarons (at T=0K and T=300K). We find a clearly decreased hopping rate in the polaron case. In the inset, we show the relative hopping strength reduction (with respect to the exciton hopping rate) over the twist angle. The polaron effect is most efficient at small twist angles with a hopping rate reduction of over 80\% at room temperature.}
    \label{fig:Hopping}
\end{figure}

\textbf{Polaron impact on hopping rates.} 
In the context of exciton transport, the hopping rate $|t|$ (cf. Eq. \eqref{Eq:Hopping_term}) emerges as a crucial factor determining the propagation of excitons. Focusing on the dominant nearest-neighbor hopping terms, we find that the hopping rate is directly dependent on the moiré exciton band structure and on the overlap of the Wannier wavefunctions (see Eq. \eqref{Eq:Hopping_term}). Plotting $|t|$ as a function of the twist angle, we find that  at small angles, the hopping is negligibly small, cf. Fig. \ref{fig:Hopping}, reflecting flat moiré exciton bands and thus trapped exciton states. As the twist angle increases, excitons gain more mobility. The question now is how this already well-known behaviour changes in presence of polarons.
We find that the formation of polarons generally results in a significant reduction of the hopping rate, cf. the solid vs dashed lines in Fig. \ref{fig:Hopping}. This can be traced back to the polaron-induced mass enhancement, impeding the motion of excitons. Notably, already at T=0K, polarons lead to a clear decrease  in the hopping rate (blue line) compared to the purely excitonic case (dashed line). As the temperature rises, exciton-phonon interaction becomes more efficient leading to a larger mass enhancement and as a direct consequence a larger hopping rate reduction (red line).  Equation \eqref{Eq:Polaron_Energies} illustrates a direct relationship between the polaron mass and the phonon occupation number. This suggests that with rising temperature, the exciton mass also increases, consequently impeding the mobility of excitons. Therefore, the phonon-mediated interaction plays a pivotal role in modifying the material's transport characteristics.
\\
Furthermore, we find a distinct angle-dependence of the polaron impact on the hopping rate. We find the absolute decrease in the rate in presence of polarons to be considerably larger at higher twist angles, cf. Fig. \ref{fig:Hopping}. There is a general increase of the exciton mobility as a function of the twist angle reflecting the decreasing effect of the moiré potential. 
Interestingly, the formation of polarons counteracts this effect by enhancing the effective exciton mass and reducing the exciton bandwidth (Figs. \ref{fig:Bandwidth}(a)-(b)), i.e. polarons move the transition from the moiré-trapped to the delocalized phase \cite{brem2020tunable} to larger twist angles. 
The inset of Fig. \ref{fig:Hopping} shows the relative polaron-induced changes in the hopping rates (compared to the free excitonic case) as a function of the twist angle for different fixed temperatures. Already at zero temperature (T=0K), the polaron effect is prominent, leading to a substantial deviation from the free exciton hopping rate due to the self-interaction via virtual phonons \cite{ortmann2009theory}. These deviations are observed to exceed 60\% at 1$^{\circ}$. This significant reduction in the hopping rate highlights the importance of polarons in altering transport properties of excitons in TMD-based materials. As the temperature rises, the contribution of thermal energy to the effective exciton mass becomes relevant via the increased phonon occupation appearing in Eq. \eqref{Eq:Polaron_Energies}. This further emphasizes the impact of polarons, resulting in even larger changes in the hopping rates. At elevated temperatures, we observe relative deviations of over 80\% at 1$^{\circ}$ from the undisturbed excitonic case. 
While the absolute rate reduction was found to increase with the twist angle (main Fig. \ref{fig:Hopping}), the relative difference is interestingly the highest at small twist angles  (inset of Fig. \ref{fig:Hopping}), as already discussed above. Specifically,  the hopping rate reduction at a twist angle of 3$^{\circ}$ is found to be around 20\% at T=0K and up to 30\% at T=300K (compared to >60 and >80\% at 1$^{\circ}$, respectively). This trend shows the intriguing role the twist angle plays for the moiré exciton transport.
\begin{figure}[t!]
    \centering
    \includegraphics[width=0.48\textwidth]{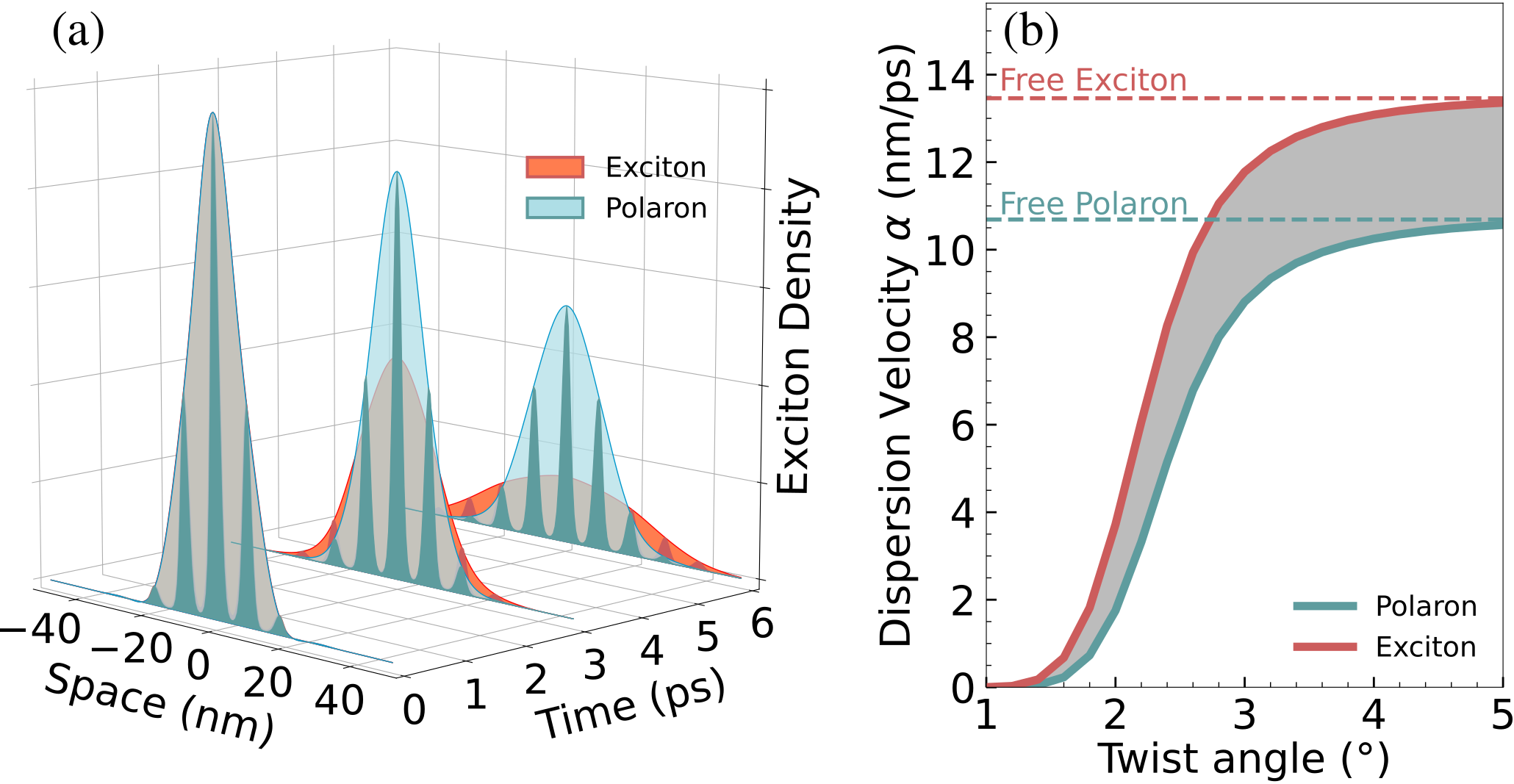}
    \caption{(a) Spatial evolution of the exciton density at 2$^\circ$  (shaded area) for both the exciton and polaron case at fixed times. Noticeably, there is a clearly slowed-down spatial propagation of polarons. (b) Dispersion parameter $\alpha$ (Eq. \eqref{Eq:alpha}) plotted against the twist angle, offering a direct comparison between the undisturbed exciton case and the polaron case. Both saturate into the free exciton/polariton case at larger twist angles, where the moiré potential becomes negligible.}
    \label{fig:alpha}
\end{figure}
Figure \ref{fig:alpha}(a) provides a visual representation of the spatial evolution of the exciton density at fixed times at a twist angle of 2$^{\circ}$ and at room temperature. Starting with an initial Gaussian distribution at time t = 0 ps, we track how excitons propagate in space. We find localized peaks (darker shading) reflecting localized states at the considered angle. In order to facilitate a clearer comparison of the propagation behaviour, we draw an envelope function and normalize the initial distributions of excitons with and without polarons to be identical (red and blue shading). At t = 3ps and more prominently at t = 6ps, we find a distinct difference in the propagation speed between free excitons and polarons (cf. Fig \ref{fig:alpha}(a)). This observation aligns with our previous findings, where polarons gave rise to a slower exciton propagation. This deceleration can be attributed to the increased effective mass of excitons due to polarons, leading to reduced hopping rates and a smaller mobility.

To quantitatively analyze the moiré exciton propagation, we utilize the dispersion velocity parameter $\alpha$ from Eq. \eqref{Eq:alpha}. The analysis reveals a clear trend: as the twist angle increases, both excitons and polarons converge toward their free excitonic/polaron solutions, respectively (Fig. \ref{fig:alpha}(b)). Beyond a certain critical angle of approx. 3$^\circ$, the dispersion parameter $\alpha$ saturates, reflecting a constant dispersion characteristic for free excitons/polarons.  Exciton propagation within a moiré potential featuring a large twist angle reflects the dispersion behavior of a quantum-mechanical wave packet, resulting in a motion reminiscent of free particle motion \cite{knorr2022exciton}. In contrast, for small twist angles, exciton propagation becomes severely limited due to band flattening in the exciton dispersion. This twist-angle effect is supported by the formation of polarons, which lead to an increase in mass and thus also to a flattening of the band. This band flattening counteracts the processes of convergence to the unperturbed case, where the excitons are delocalised. As a result, the polaronic dispersion velocity is lower than in the purely excitonic case. Moreover, we observe a delay in the velocity increase (grey area in Fig \ref{fig:alpha}). The relative difference in dispersion velocity between free excitons and polarons at large twist angles can be determined to be larger than 50\%. 

\section{Summary}
In this study, we have investigated the influence of the polaron-induced exciton mass enhancement on the moiré exciton transport in a twisted MoSe$_2$/WSe$_2$ heterostructure. The formation of polarons leads to a distinct flattening of  moiré exciton bands and to a significant reduction in their bandwidth. At small twist angles, changes in the energy bandwidth are predicted to exceed 60\%. We have also examined the polaron-induced change in the hopping rates, which is a key quantity for exciton propagation. At small angles, the hopping strength was significantly diminished by up to 80\%  due to polaron-induced band flattening and thus weaker nearest-neighbour wavefunction overlap. Generally, at  higher twist angles, exciton mobility improves owing to the reduced confinement within the moiré potential. However, polarons are found to counteract this effect by quenching the exciton's zero-point confinement energy, impeding exciton motion in TMD heterostructures. Our findings contribute to a better microscopic understanding of moiré exciton propagation in presence of polarons.

\section*{Acknowledgements}
We acknowledge funding from the Deutsche Forschungsgemeinschaft
(DFG) via SFB 1083.

\bibliography{Bib}
\end{document}


\title{Polaron-induced changes in moiré exciton propagation in twisted van der Waals heterostructures\\
\Large Supplementary Information}
\author{Willy Knorr}
\email{knorrw@uni-marburg.de}
\author{Samuel Brem}
\author{Giuseppe Meneghini}
\author{Ermin Malic}
\affiliation{Department of Physics, Philipps University, 35037 Marburg, Germany}
\date{\today}
\maketitle
\linespread{1}
\subsection{Effective Single Particle Hamiltonian}
In our pursuit of evaluating the excitonic band structure within the presence of a moiré potential, we derive an effective single-particle Hamiltonian. In order to accomplish this, we introduce operators for electron-hole pairs, represented as $P^{\dagger}_{l\mathbf{k},l'\mathbf{k}'}$. These operators are defined as a summation over excitonic eigenmodes where $l$ represents the layer index \cite{brem2020tunable}. These modes incorporate factors $\alpha_{ll'}/\beta_{ll'}$, accounting for the electron and hole masses in the conduction and valence bands, respectively. For a WSe$_2$-MoSe$_2$ bilayer, these values are determined as $m_e/m_0=0.36(0.6)$ and $m_h/m_0=0.29(0.5)$ based on Ref. \cite{kormanyos2015k}. The exciton wave functions satisfy the Wannier equation \cite{koch2006semiconductor}. In the low-density regime, with a focus on the excitonic ground state, we arrive at the free exciton Hamiltonian \cite{brem2020tunable}.
\begin{equation}
    H_{ex} = \sum_{ll'\textbf{Q}}\varepsilon_{\textbf{Q}}^{ll'}X^{\dagger}_{ll'\textbf{Q}}X_{ll'\textbf{Q}}.
\end{equation}
where $\varepsilon_{\textbf{Q}}^{ll'}$ denotes the free exciton dispersion. Additionally, we transform the Hamiltonian for the moiré potential \cite{brem2020tunable} reading
\begin{align}
H_M&=\sum_{ll'\textbf{Q}\textbf{q}}\text{M}_{\textbf{q}}^{ll'}X^{\dagger}_{ll'\textbf{Q}+\textbf{q}}X_{ll'\textbf{Q}}+h.c.
\end{align}
with the matrix element $\text{M}_{\textbf{q}}^{ll'} = V_l^c(\textbf{q})\mathcal{J}_{ll'}(\beta_{ll'}\textbf{q})-V_{l'}^v(\textbf{q})\mathcal{J}^*_{ll'}(\alpha_{ll'}\textbf{q})$ including the electronic moiré potential $V^{v/c}_l$ in the valence and conduction band respectively, as well as the excitonic form factor $\mathcal{J}_{ll'}(\textbf{q})=\sum_k\psi^*_{ll'}(\textbf{k})\psi_{ll'}(\textbf{k}+\textbf{q})$ \cite{brem2020tunable}. For the electronic moir\'e potentials we employ the microscopic approach developed in Ref. \cite{brem2020tunable} yielding
\begin{align}
    H_M &= \sum_{L\textbf{Q}n}\zeta_LX^{\dagger}_{L\textbf{Q}+(-1)^{l_e}\textbf{G}_n}X_{L\textbf{Q}}\quad\text{with}\\
    \zeta_L&=\begin{cases}v_{l_e}^c\mathcal{J}_{L}(\beta_L\textbf{G}_0)-v_{l_h}^v\mathcal{J}_{L}(\alpha_L\textbf{G}_0)~~\text{for}~l_e=l_h\\ v_{l_e}^c\mathcal{J}_{L}(\beta_L\textbf{G}_0)-v_{l_h}^{v*}\mathcal{J}_{L}(\alpha_L\textbf{G}_0)~~\text{for}~l_e\neq l_h\end{cases},
\end{align}
where $L=(l_e,l_h)$ and the parameters $v_{l}^\lambda$ are extracted from first principle computations and read $v_{l}^\lambda = \alpha_l+exp(2\pi i\sigma/3)\beta_l$ \cite{brem2020tunable}. In this equation $\alpha/\beta$ are deduced moiré parameter which can be found in the table below \cite{brem2020tunable}, the parameter $\sigma$ represents the stacking parameter. Throughout this work we consider R-type stacking ($\sigma = 1$) where both layers have a parallel orientation. 
\begin{table}[h]
    \centering
    \caption{Required DFT input parameters for our microscopic model. }
    \begin{tabular}{|c|c|c|}
        \hline
        band  &  $\alpha$[meV] & $\beta$[meV] \\
        \hline
        vb-1(Mo) & -20 & -7.3 \\
        vb(W) & -16.1 & -8\\
        cb(Mo) & -14.7 & -5.2 \\
        cb+1(W) & -11.1 & -5.4\\
        \hline
    \end{tabular}
    
    \label{tab:my_label}
\end{table}
The moiré potential creates a superlattice with reciprocal lattice vectors $\mathbf{G_n}=C_3^n \mathbf{G_0}$, which only allows the mixing between discrete center-of-mass momenta. We utilise a zone-folding approach to take advantage of the new periodic lattice \cite{hagel2021exciton}. Hence, by changing into the in the zone-folded eigenbasis $F^{\dagger}_{Ls\mathbf{Q}}=\tilde{X}_{L,\mathbf{Q}+\mathbf{G}_s}$ we obtain
 \begin{align}
     H = \sum_{Ls\textbf{Q}}\tilde{\varepsilon}_{L,\textbf{Q}+s_1\textbf{G}_1+s_2\textbf{G}_2}F^{\dagger}_{Ls\textbf{Q}}F_{Ls\textbf{Q}}+\sum_{Lss'\textbf{Q}}\tilde{\mathcal{M}}_{ss'}^{L}F^{\dagger}_{Ls\textbf{Q}}F_{Ls'\textbf{Q}}
 \end{align}
 with the modified moiré mixing matrix  
 \begin{align}
     \tilde{\text{M}}_{ss'}^{L} &= \zeta_L\Bigg(\delta(s_1,s_1'+(-1)^{l_e})\delta(s_2,s'_2)+\delta(s_1,s_1')\delta(s_2,s'_2+(-1)^{l_e})+\delta(s_1,s'_1-(-1)^{l_e})\delta(s_2,s'_2-(-1)^{l_e})\Bigg)\notag\\
     &+\zeta^*_L\bigg(\delta(s_1,s_1'-(-1)^{l_e})\delta(s_2,s'_2)+\delta(s_1,s_1')\delta(s_2,s'_2-(-1)^{l_e})+\delta(s_1,s'_1+(-1)^{l_e})\delta(s_2,s'_2+(-1)^{l_e})\bigg).
 \end{align}
We change to the eigenbasis
 \begin{align}
     Y^{\dagger}_{L\nu\textbf{Q}} = \sum_sc_{Ls}^{\nu*}(\textbf{Q})F^{\dagger}_{Ls\textbf{Q}}.
 \end{align}
 where the coefficients fulfil the eigenvalue equation
 \begin{align}
     \tilde{\varepsilon}_{L,\textbf{Q}+s_1\textbf{G}_1+s_2\textbf{G}_2}c_{Ls}^{\nu}(\textbf{Q})+\sum_{s'}\tilde{\text{M}}_{ss'}^{L}C_{Ls'}^{\nu}(\textbf{Q}) = E_{L\mu\textbf{Q}}C_{Ls}^{\nu}(\textbf{Q}).
 \end{align}
This new basis leads us to the final diagonal Hamiltonian 
 \begin{align}\label{Eq:singleHamiltonianFinal}
     H = \sum_{L\mu\textbf{Q}}E_{L\mu\textbf{Q}}Y_{L\mu\textbf{Q}}^{\dagger}Y_{L\mu\textbf{Q}}
 \end{align}
 which is further discussed in the main text.

\subsection{Exicton-Phonon Interaction}
We introduce the electron-phonon interaction Hamiltonian. We simplify the treatment of phonons by using a Taylor expansion near high symmetry points, incorporating material-specific phonon energies. The phonon dispersion, akin to electronic band structure, is characteristic to each material and can be determined through first principles calculations. These DFPT results are incorporated into the electron-phonon matrix element $D_{\textbf{q}j}^{\lambda}$ \cite{jin2014intrinsic,brem2020phonon}. The Hamiltonian reads
\begin{align}
    H_{\text{el-ph}}=\sum_{\lambda,\textbf{q},\textbf{k},j}D_{\textbf{q}j}^{\lambda} a^{\dagger}_{\lambda,\textbf{k}+\textbf{q}}a_{\lambda,\textbf{k}}\Bigg(b_{\textbf{q}}^j+b_{-\textbf{q}}^{\dagger j}\Bigg),
\end{align}
with $\textbf{k}$ and $\textbf{q}$ denoting electron and phonon momenta, respectively. Here, $\lambda$ signifies the band index, and $j$ represents the phonon mode index. In this context, the operator $a_{\lambda,\textbf{k}}$ corresponds to the electron operator, while $b_{\textbf{q}}^j$ stands for the phonon operator.\\
By employing again electron-hole pair operator introduced in the previous section we perform a basis transformation into an excitonic basis \cite{christiansen2019theory}. We obtain the following Hamiltonian
\begin{align}
    H_{\text{ex-ph}}=\sum_{\textbf{Q},\textbf{q},j,\mu,\nu}\mathcal{D}_{\textbf{q}j}^{\mu\nu} X^{\dagger}_{\mu,\textbf{Q}+\textbf{q}}X_{\nu,\textbf{Q}}\Bigg(b_{\textbf{q}}^j+b_{-\textbf{q}}^{\dagger j}\Bigg)
\end{align}
with the transformed exciton-phonon matrix 
\begin{align}
    \mathcal{D}^{\mu\nu}_{\textbf{q}j} = D^c_{\textbf{q}j}\mathcal{J}^{\mu\nu}(\beta\textbf{q})-D^c_{\textbf{q}j}\mathcal{J}^{\mu\nu}(-\alpha\textbf{q}).
\end{align}
Here once again $\mathcal{J}^{\mu\nu}(-\alpha\textbf{q})$ denotes the excitonic form factor. 

\subsection{Polaron Transformation}
Utilizing the Hamiltonian introduced earlier, we can construct a comprehensive Hamiltonian that encapsulates the entirety of our system's dynamics and interactions. This encompassing Hamiltonian enables us to thoroughly describe and analyze the behavior and properties of our system, taking into account both the free excitonic contributions denoted by $H_{\text{ex}}$ and the phonon-related aspects characterized by $H_{\text{ph}}$. Please note that for the sake of simplicity, we omit the phonon mode index $j$. It's important to emphasize that while we do not explicitly write out this index, it still exists and is applicable in our calculations. Furthermore, it incorporates the critical interaction term, $H_{\text{ex-ph}}$, which embodies the intricate interplay between excitons and phonons, allowing us to gain a deeper understanding of the intricate phenomena occurring within our system.
\begin{align}
    H =  H_{\text{ex}} + H_{\text{ph}} + H_{\text{ex-ph}} = H_0 + H_{\text{ex-ph}},
\end{align}
We employ a polaron transformation to facilitate a unitary mapping of Hamiltonians.
\begin{align*}
    \widetilde{H} &= e^{-S}He^{S}\\
    &= H + [S,H] + \frac{1}{2}\Big[S,[S,H]\Big] + \mathcal{O}(H^3)\\
    &= H_0 + H_{\text{ex-ph}} + [S,H_0] + [S,H_{\text{ex-ph}}] + \frac{1}{2}\Big[S,[S,H]\Big] + \mathcal{O}(H^3)
\end{align*}
We omit the above Baker-Campbell-Hausdorff in third order. We define the transformation operator  
\begin{align}
    S = -\sum_{\textbf{Q}\textbf{q}}\mathcal{D}_{\textbf{q}}\Bigg(\frac{1}{\mathcal{E}_{\textbf{Q}+\textbf{q}}-\mathcal{E}_{\textbf{Q}}+\hbar\Omega}b_{-\textbf{q}}^{\dagger}+\frac{1}{\mathcal{E}_{\textbf{Q}+\textbf{q}}-\mathcal{E}_{\textbf{Q}}-\hbar\Omega}b_{\textbf{q}}\Bigg)X_{\textbf{Q}+\textbf{q}}^{\dagger}X_{\textbf{Q}}.
\end{align}
As one can easily ascertain, the commutator relation below follows from this transformation
\begin{align}
    [S,H_0] = -H_{\text{ex-ph}}.\label{Eq:ComH0}
\end{align}
Thus, we obtain the new Hamiltonian
\begin{align}
    \widetilde{H} = H_0 +\frac{1}{2}[S,H_{\text{ex-ph}}].
\end{align}
Using many-particle Fock states of the unperturbed Hamiltonian, denoted as $\ket{n}$ with corresponding eigenvalues $E_n$, we can derive the matrix elements of the operator S in this eigenbasis from Equation \eqref{Eq:ComH0}. 
\begin{align}
    \bra{n}S\ket{m}= \frac{\bra{n}H_{\text{ex-ph}}\ket{m}}{E_n-E_m}
\end{align}
Which yields for our new Hamiltonian \cite{kittel1987quantum,czycholl2015theoretische}
\begin{align}
    \widetilde{H} = H_0 - \frac{1}{2}\sum_{lmn}\bra{l}H_{\text{ex-ph}}\ket{m}\bra{m}H_{\text{ex-ph}}\ket{n}\Bigg(\frac{1}{E_m-E_n}-\frac{1}{E_l-E_m}\Bigg)\ket{l}\bra{n}
\end{align}
In order to obtain an effective Hamiltonian, we perform a trace over the phonons within a bath approximation. As a consequence, the initial states represented as $\ket{n}$ and the final states represented as $\ket{l}$ must maintain the same phonon configuration. This constrains the processes to involve a two-step sequence where a phonon is first absorbed and then subsequently emitted with the same momentum or vice versa. Additionally, we need to account for energy conservation in these processes. As an example, considering one of these processes, the energy change associated with the absorption of a phonon with energy $\hbar\Omega$ can be expressed as $E_m - E_n = \mathcal{E}_{\textbf{Q}+\textbf{q}} - \mathcal{E}_{\textbf{Q}}-\hbar\Omega$. It's important to note that we intentionally neglect two-particle processes in our calculations. This set of assumptions and considerations ultimately leads us to the final form of the Hamiltonian.
\begin{align*}
    \widetilde{H} &= H_0 - \sum_{\mathbf{Q},\mathbf{q}}|\mathcal{D}_{\mathbf{q}}|^2\left(\frac{n_q}{\Delta \mathcal{E}-\hbar\Omega}+\frac{n_{q}+1}{\Delta \mathcal{E}+\hbar\Omega}\right)X_{\mathbf{Q}}^{\dagger}X_{\mathbf{Q}}.
\end{align*}
where $\Delta \mathcal{E}=\mathcal{E}_{\textbf{Q}+\textbf{q}}-\mathcal{E}_{\textbf{Q}}$. We can identify a polaron renormalized energy as 
\begin{align*}
    \widetilde{\mathcal{E}}_{\mathbf{Q}} &= \mathcal{E}_{\mathbf{Q}} - \sum_{\mathbf{q}}|\mathcal{D}_{\mathbf{q}}|^2\left(\frac{n_q}{\Delta \mathcal{E}-\hbar\Omega}+\frac{n_{q}+1}{\Delta \mathcal{E}+\hbar\Omega}\right).
\end{align*}
Performing a Taylor expansion on the second term for small \textbf{Q} we obtain
\begin{align*}
    \widetilde{\mathcal{E}}_{\mathbf{Q}} = \mathcal{E}_{\mathbf{Q}} - \sum_{\mathbf{q}}|\mathcal{D}_{\mathbf{q}}&|^2\Bigg[\frac{n_q}{\mathcal{E}(\textbf{q})-\hbar\Omega}\Bigg\{1-\frac{\frac{\hbar^2}{m}\textbf{Q}\cdot\textbf{q}}{(\mathcal{E}(\textbf{q})-\hbar\Omega)}+\Bigg(\frac{\frac{\hbar^2}{m}\textbf{Q}\cdot\textbf{q}}{(\mathcal{E}(\textbf{q})-\hbar\Omega)}\Bigg)^2\Bigg\}\\
    &+\frac{(n_{q}+1)}{\mathcal{E}(\textbf{q})+\hbar\Omega}\Bigg\{1-\frac{\frac{\hbar^2}{m}\textbf{Q}\cdot\textbf{q}}{(\mathcal{E}(\textbf{q})+\hbar\Omega)}+\Bigg(\frac{\frac{\hbar^2}{m}\textbf{Q}\cdot\textbf{q}}{(\mathcal{E}(\textbf{q})+\hbar\Omega)}\Bigg)^2\Bigg\}\Bigg]
\end{align*}
Upon integration over $\textbf{q}$, the contributions linear in $\textbf{q}$ vanish. Consequently, we are left with two key terms: one responsible for shifting the original exciton energy, denoted as $\mathcal{E}_{\text{Polaron}}$ (representing the polaron shift), and the other term corresponds to a mass renormalization factor denoted as $\lambda$. In a more concise form, we can express the new energy as:
\begin{align}
    \widetilde{\mathcal{E}}_\mathbf{Q} &= -\mathcal{E}_{\text{Polaron}} + \frac{\hbar^2\mathbf{Q}^2}{2m^*},
\end{align}
with $m^* = (1+\lambda)m$. The individual terms can be specified as follows:
\begin{align}
    \mathcal{E}_{\text{Polaron}} = &\sum_{\textbf{q}} |\mathcal{D}_{\mathbf{q}}|^2 \Bigg( \frac{n_q}{\mathcal{E}(\textbf{q})-\hbar\Omega}+\frac{(n_{\mathbf{q}}+1)}{\mathcal{E}(\textbf{q})+\hbar\Omega}\Bigg)\\
    \lambda = \frac{2\hbar^2}{m} &\sum_{\textbf{q}} |\mathcal{D}_{\mathbf{q}}|^2 \textbf{q}^2\Bigg(\cancelto{0}{ \frac{n_q}{(\mathcal{E}(\textbf{q})-\hbar\Omega)^3}}+\frac{(n_{\mathbf{q}}+1)}{(\mathcal{E}(\textbf{q})+\hbar\Omega)^3}\Bigg).
\end{align}
Observing the mass renormalization term, it becomes apparent that the first term vanishes during the integration over $\textbf{q}$. This can be confirmed through a residue analysis.

\bibliography{Bib}